# Diameter-Dependent Electron Mobility of InAs Nanowires

Alexandra C. Ford[1,2,3,†], Johnny C. Ho[1,2,3,†], Yu-Lun Chueh[1,2,3,†], Yu-Chih Tseng[1], Zhiyong Fan[1,2,3], Jing Guo[4], Jeffrey Bokor[1,2], Ali Javey[1,2,3,]*

[1]Department of Electrical Engineering and Computer Sciences, University of California at Berkeley, Berkeley, CA, 94720, USA.

[2]Materials Sciences Division, Lawrence Berkeley National Laboratory, Berkeley, CA 94720, USA.

[3]Berkeley Sensor and Actuator Center, University of California at Berkeley, Berkeley, CA, 94720, USA.

[4]Department of Electrical and Computer Engineering, University of Florida, Gainesville, Florida 32611

[†] These authors contributed equally.

* Corresponding author: ajavey@eecs.berkeley.edu

ABSTRACT – Temperature-dependent *I-V* and *C-V* spectroscopy of single InAs nanowire field-effect transistors were utilized to directly shed light on the intrinsic electron transport properties as a function of nanowire radius. From *C-V* characterizations, the densities of thermally-activated fixed charges and trap states on the surface of untreated (i.e., without any surface functionalization) nanowires are investigated while enabling the accurate measurement of the gate oxide capacitance; therefore, leading to the direct assessment of the field-effect mobility for electrons. The field-effect mobility is found to monotonically decrease as the radius is reduced to sub-10 nm, with the low temperature transport data clearly highlighting the drastic impact of the surface roughness scattering on the mobility degradation for miniaturized nanowires. More generally, the approach presented here may serve as a versatile and powerful platform for in-depth characterization of nanoscale, electronic materials.

Semiconductor nanowires (NWs) have tremendous potential for applications in high performance nanoelectronics and large-area flexible electronics.[1,2,3,4,5,6,7,8,9] In particular, InAs NWs are promising as the channel material for high performance transistors because of their high electron mobility and ease of near-ohmic metal contact formation.[10,11,12,13,14] Of particular interest is the dependence of the carrier mobility on NW radius for a given material, especially since smaller NWs are highly attractive for the channel material of nanoscale transistors as they enable improved electrostatics and lower leakage currents. Most theoretical studies have found carrier mobility to increase with radius for sub-10 nm Si NWs (no data available for InAs NWs), either attributing the trend to the dominant surface roughness scattering in smaller radius NWs,[15] or an enhanced phonon scattering rate due to an increased electron-phonon wavefunction overlap in smaller radius NWs.[16,17] On the other hand, experimental reports in the literature have been contradictory, ranging from observation of mobility enhancement[18] to degradation[19] with Si nanowire miniaturization for diameters (or widths) down to 10 nm. Therefore, the diameter dependency of the mobility highly depends on the specific nanowire material system,[15,16,17,18,19,20] the diameter range, and the method used to assess the electron mobility. The challenge in attaining accurate experimental data mainly arises from the difficulty of ohmic contact formation to nanoscale materials and the direct measurement of the gate capacitance. Here, we report the detailed current-voltage (*I-V*) and capacitance-voltage (*C-V*) spectroscopy of individual InAs NWs with ohmic contacts at different temperatures; therefore, enabling the direct assessment of field-effect mobility as a function of NW diameter while elucidating the role of surface/interface fixed charges and trap states on the electrical properties.

InAs NWs used in this study were synthesized on Si/SiO$_2$ substrates by physical vapor transport method using Ni nanoparticles as the catalyst as previous reported.[21] The grown InAs

NWs were over 10 µm long with a radius range of 7-20 nm (Fig. 1a). The NWs are single crystalline with a native oxide thickness of 2-2.5 nm as evident from TEM analysis (Figs. 1b and c). Importantly, the NWs grown using our previously reported conditions do not exhibit any noticeable tapering effect with a uniform diameter along the length of each NW, as confirmed by TEM and SEM. Energy dispersion spectrometry (EDS), as shown in Figure 1d, indicates that the chemical composition of In:As is close to 1:1.

For the electrical transport measurements, field-effect transistors in a back-gated configuration were fabricated (Figs. 1e-f, also see Supp. Info. for fabrication details). Electrical properties of representative FETs with NW radius $r$=7.5-17.5 nm are shown in Figure 2. Long channel lengths, $L$=6-10 µm, were used for this study in order to ensure diffusive transport of carriers (rather than ballistic or quasi-ballistic transport), from which intrinsic transport properties, such as carrier mobility, can be deduced. Although the NWs were not intentionally doped, as expected, the devices exhibit an n-type behavior due to the high electron concentration of "intrinsic" InAs, arising from surface fixed charges and possible local imbalance in stoichiometry. Notably, for the channel lengths and NW diameters explored in this study, a linear dependence of the device resistance as a function of channel length is observed which is indicative of ohmic metal source/drain contacts (Ni) to InAs NWs.[22] From the $I$-$V$ characteristics (Fig. 2b-d), it is clearly evident that larger diameter NWs exhibit higher ON currents and more negative threshold voltages. Specifically, unit length normalized ON currents ($V_{DS}$=2 V and $V_{GS}$-$V_t$=6 V) of ~ 40, 110, and 140 µA-µm are obtained for $r$=7.5, 12.5, and 17.5 nm NWs, respectively. This trend can be attributed to a larger cross-sectional area (i.e., effective channel width) for large diameter NWs, but could also be indicative of reduced carrier scattering with increasing diameter. To further shed light on this trend, investigation of the electron transport properties as a function of NW radius is imperative. Particularly, electron mobility, $\mu_n$, is an important figure of merit as it relates the drift velocity of electrons to an applied electric-field. Accurate and direct measurement of the gate oxide capacitance, however, is needed for the extraction of field-effect mobility from *I-V* characteristics.

In order to determine the gate oxide capacitance of NW FETs, and to shed light on the density and characteristics of the surface/interface trap states and fixed charges, direct *C-V* measurements were performed on *single*-InAs NW devices at various temperatures. Previously, the only reported *C-V* measurements for InAs NW FETs have been for parallel arrays of NWs (>100 vertical NWs per device) and at room temperature.[23] For the purpose of this work, temperature-dependent *C-V* spectroscopy of *single*-NW devices with known NW radius are required to minimize the averaging effects and shed light on properties of individual NWs. Hence, we utilized a method previously developed by Ilani, S., et al. in order to measure the small capacitance signal (10aF-1fF) of *single*-NW FETs over a large background parasitic capacitance (~30fF).[24,25] A similar method was also utilized in the past by Tu, R., et al. to examine the gate oxide capacitance of single Ge NW-FETs.[25] As depicted in Figure 3a, buried-gate InAs NW FETs with $t_{ox}$~60 nm, S/D length $L_{SD}$~10 µm, and buried-gate length $L_{LG}$~5 µm were fabricated. Details of the *C-V* measurement set-up and device fabrication can be found in the Supporting Information.

Figure 3b shows the temperature dependency of *C-V* characteristics for a representative InAs NW-FET ($r$~11nm, $L_G$=4.7µm, $L_{SD}$=9.3µm) obtained with an AC signal of 125mV at 2kHz. For this device, a flat-band voltage of $V_{FB}$~0V (corresponding to the on-set voltage of the sharp decrease in the measured capacitance) is observed, with $V_{LG}>V_{FB}$~0V resulting in the accumulation of electrons in the n-type InAs channel (i.e., ON state). We note that this is in distinct contrast to the operation mode of the conventional MOSFETs in which the ON state corresponds to the inversion of the channel (rather than accumulation). The gate capacitance value obtained in the accumulation regime corresponds to the oxide capacitance, $C_{LG,accumulation}=C_{ox}$, which is temperature independent. When $V_{LG} < V_{FB}$ (i.e., $V_{LG}<0$ V), the channel is depleted of electrons, thus resulting in the reduction of the total gate capacitance due to the addition of the semiconductor capacitance, $C_s$, in series with $C_{ox}$ (i.e., $C_{LG,depletion} = C_{ox} C_s / (C_{ox}+C_s)$). At this state, the NW channel is effectively turned "OFF". The temperature dependent $C$-$V$ measurements illustrate two important effects (Fig. 3b). First, a shift in $V_{FB}$ is observed as a function of temperature which can be attributed to the change in the population density of the thermally activated, donor-like fixed charges, $N_s$ (near the conduction band edge, in the unit of /cm$^2$), at the NW surface/interface. Second, the capacitance in the depletion region is drastically reduced as the temperature is lowered from 200K to 150K, but relatively unchanged thereafter. This trend is a clear signature of thermally activated, surface/interface traps (with density, $D_{it}$) as they induce a capacitance, $C_{it}$, in parallel to $C_s$ (Fig. 3a); therefore, effectively increasing $C_{LG,depletion}$. For this case, the gate capacitance in the depletion regime is given as, $C_{LG,depletion} = C_{ox}(C_s+C_{it}) / (C_{ox}+C_s+C_{it})$. Below 150K, the measured depletion capacitance is independent of temperature, indicating that the traps stop responding. Based on this analysis, we extrapolate a $C_s$~10.5 aF and $C_{it}$~0, 11.3, 316 aF at 77, 150, 200K, respectively. Similar $C_{it}$ values with $D_{it}$~2x10$^{11}$ states cm$^{-2}$eV$^{-1}$ at 200K were obtained from frequency-dependent measurements (2 and 20kHz, see Supporting Information, Figs. S1 and S2). It is important to note that 2kHz may not present the true low frequency operation regime as some traps may already be irresponsive at that frequency. Therefore, the extracted $D_{it}$ values only represent a lower bound limit. We were not able to perform $C$-$V$ measurements at temperatures higher than 200K due to the thermal noise and leakage currents of low band-gap ($E_g$~0.36 eV) InAs NW channels (arising from the band-to-band thermal generation of carriers).

Detailed electrostatic modeling was also performed to further investigate the effect of fixed charges and trap states on the *C-V* characteristics. A two-dimensional Poisson equation was self-consistently solved with the equilibrium carrier statistics for the InAs NW and the native oxide layer for a cross section perpendicular to the nanowire axis. Both $N_S$ and $D_{it}$ are treated as the fitting parameters in the simulation. A close fit of the experimental data for the normalized gate capacitance, as shown in Fig. 3b, is obtained when assuming $N_s$=0, 1.5x10$^{11}$, 4.5x10$^{11}$ states cm$^{-2}$ and $C_{it}$=0, 17.4, 344 aF for 77, 150, 200 K, respectively, which is consistent with the values extrapolated from the analytical expressions described above. When quantum effects[23] are taken into consideration by self-consistently solving the Poisson and Schrödinger equations in the quantum simulation, it is found that quantum effects decrease the semiconductor capacitance by shifting the centroid of the charge away from the NW surface. However, because in our fabricated FETs, the gate oxide thickness is much larger than the NW radius (~3 to 7 times larger), the quantum effects on the total gate capacitance are relatively small (Supporting Information, Fig. S4).[26]

In addition to the detailed characterization of $C_{it}$ and $D_{it}$, we were able to directly measure $C_{ox}$ as a function of NW radius. Figure 4 shows the experimentally obtained $C_{ox}$ for different NW-FETs with *r*=10-20 nm. We also performed electrostatic modeling of the oxide capacitance values by using the finite element analysis software package Finite Element Method Magnetics (Fig. 4). The measured and modeled capacitance values are in qualitative agreement with the experimental values ~25% higher than the modeled results. We attribute this discrepancy to the infringing capacitances between LG and the underlapped NW segments which were ignored in the simulation as well as the geometric uncertainties associated with the fabricated NW-FETs (i.e., the exact thickness of the gate oxide deposited on Pt LGs). Additionally, $C_{ox}$ was calculated from the analytical expression, $C_{ox} = \dfrac{2\pi\varepsilon\varepsilon_0 L}{\cosh^{-1}[(r+t_{ox})/r]}$, which corresponds to the capacitance of a cylindrical wire on a planar substrate and is often used in the literature for NW device analysis.[26,27,28,29,30] Here, $\varepsilon$ is the dielectric constant of the gate insulator ($\varepsilon$=3.9 for SiO$_2$) and $\varepsilon_0$ is the permittivity of free space. The capacitance values obtained from this analytical expression are ~2x higher than the experimental values (Fig. 4), demonstrating the lack of accuracy of this analytical method for NW-FET performance analyses.

From the *C-V* and *I-V* measurements, we next assess the field-effect electron mobility of InAs NW FETs by using the low-bias ($V_{DS}$=0.1 V) transconductance, $g_m = \left.\dfrac{dI_{DS}}{dV_{GS}}\right|_{V_{DS}}$, and the analytical expression, $\mu_n = g_m \times \dfrac{L^2}{C_{ox}} \times \dfrac{1}{V_{DS}}$. Figure 5a shows $\mu_n$ as a function of $V_{GS}$ for three different NW radii, corresponding to the $I_{DS}$-$V_{GS}$ plot of Figure 2a. It is clearly evident that the peak field-effect mobility is enhanced for larger diameter NWs with $\mu_n$~2,500, 4,000, and 6,000 cm$^2$/Vs for $r$~7.5, 12.5, and 17.5 nm, respectively. Notably, the $\mu_n$-$V_{GS}$ characteristics for all measured NW-FETs exhibit a near identical behavior with the field-effect mobility at first increasing with $V_{GS}$-$V_t$ before sharply decaying at high electric fields. This decay can be attributed to the enhanced surface scattering of the electrons at high gate fields, similar to the behavior that is observed in conventional Si MOSFETs. In addition, in quasi 1-dimensional (1-D) NWs, due to the quantization of sub-bands, the metal contacts may not enable a sufficient injection of electrons into the channel at high electric-fields as desired by the gate potential. Because of the finite sub-band energy spacing, Schottky barriers to the higher sub-bands are often formed at the NW-metal contact interfaces, therefore, lowering the transconductance and the mobility of the FETs at high gate voltages. While at a first glance, the NWs used in this study may seem rather large to exhibit quantization effects, due to the large Bohr radius of InAs (~34 nm),[13] even a $r$=10 nm NW can be treated as quasi 1-D because the confinement energies for the lowest and 2$^{nd}$ lowest sub-bands are ~100 and 240 meV, respectively (Supporting Information, Fig. S3).[31] Figure 5b illustrates the peak field-effect mobility as a function of InAs NW radius for more than 50 different FETs with $r$=7-18 nm. Over this NW radius range, the peak mobility linearly increases with radius with a slope of ~422 (cm$^2$/Vs)/nm. We note that larger or smaller radii beyond the range reported here were not explored due to the difficulty with their growth using our condition.[21] The linear drop in the field-effect mobility with reduced NW radius may be attributed to a number of factors, including the enhanced phonon-electron wavefunction overlap (i.e., enhanced phonon scattering of electrons), the increased surface scattering, enhanced defect scattering, and the lower effective gate coupling factor due to the surface states ($D_{it}$)[11] for miniaturized NWs with high surface area to volume ratio. Additionally, at a first glance, a diameter dependent contact resistance may be expected which could also affect the extracted field-effect mobility.[32] However, that appears not to be the case since for the diameter and length regime explored in this study, we find a linear dependence of the ON-state resistance as a function of the channel length.[22] Therefore, the main source of the total device resistance is due to the channel resistance.

It should be noted that the electron mobility reported in this work is the so-called "field-effect" mobility, distinct from the effective mobility and the Hall mobility. The Hall mobility represents the bulk carrier transport with no major contributions from the surface and quantization effects while both the field-effect and effective mobilities are used to characterize the carrier transport in the surface inversion (or accumulation, in the case of InAs NWs) layer of the MOSFETs. The field-effect and effective mobilities are, however, deduced from the *I-V* characteristics by using different analytical models. Specifically, the effective mobility is deduced from the drain conductance, $g_D = \left. \frac{dI_{DS}}{dV_{DS}} \right|_{V_{GS}}$ with $\mu_{n,eff} = g_D \times \frac{L^2}{C_{ox}} \times \frac{1}{(V_{GS} - V_t)}$. On the other hand, as described above, the field-effect mobility is deduced from the transconductance, $g_m$. Therefore, the main difference between the field-effect and effective mobility is the neglect of the gate electric-field dependence in the field-effect mobility expression.[33] For the device modeling, effective mobility is often used to predict the current and switching speeds. A difficulty in the accurate assessment of the effective mobility arises from the error associated with finding $V_t$ from the measured *I-V* characteristics. Therefore, for the purpose of this work, we focus on the presentation of the field-effect mobility which presents the lower bound value of the true electron mobility in InAs NWs (field-effect mobility is lower than the effective mobility, except for low gate fields). However, even if the effective mobility analysis is used, a similar diameter dependency for the peak mobility is observed for InAs NWs as depicted in Fig. S5 of the Supporting Information.

In an effort to shed light on the source of mobility degradation for smaller NWs, temperature-dependent electron transport measurements were conducted. Typical $I_{DS}$-$V_{GS}$ plots at $V_{DS}$=0.01 V for a back-gated NW device with *r*=18 nm and *L*=6.7 μm are shown in Fig. 6a over a temperature range of 50-298 K. Figure 6b shows the corresponding peak field-effect mobility as a function of temperature for this device, showing a linear enhancement of the peak electron field-effect mobility from ~6,000 to 16,000 cm$^2$/Vs as the temperature is reduced from 298 K to 200 K. Below ~200K, minimal change in the field-effect mobility is observed. We attribute this to the transition temperature at which the surface roughness scattering becomes dominant over other scattering events caused by acoustic phonon and/or surface/interface trap states. Additionally, at lower temperatures, since the surface trap states are fully frozen, they should not have an impact on the gate coupling factor. The dependency of field-effect mobility on the NW radius was also investigated at different temperatures and the data for four NW FETs with $r$=8-20 nm at 298K and 50K is shown in Fig. 6c. Even at low temperatures (i.e, 50 K), in the regime where phonons and surface/interface traps are frozen out, the monotonic increase of mobility with radius is clearly evident. Specifically, at 50 K, a near-linear trend is observed for small radius NWs (i.e., $r \leq 12$ nm, slope of ~2077 (cm$^2$/Vs)/nm) with the field-effect mobility approaching a saturation value of ~18,000 cm$^2$/Vs for larger NWs (i.e., $r$>18 nm). The phonon population is drastically reduced at 50 K, and therefore, the acoustic phonon scattering for low-field transport can be assumed to be non-existent. Additionally, most surface/interface traps are frozen out at such low temperatures and should not affect the gate electrostatic coupling or the electron transport properties near the surface. Impurity scattering should not be a factor since the NWs are not intentionally doped. As a result, the observed dependency of electron field-effect mobility on NW radius at 50K is mainly attributed to the enhanced surface roughness scattering of electrons in the miniaturized NWs. As the NW radius is reduced, electron transport near the surface dominates the electrical characteristics. However, the atomic roughness of the surface results in an enhanced carrier scattering, therefore, effectively lowering the carrier mobility. Specifically, the surface roughness scattering rate depends on the surface-area to volume ratio; therefore, a near linear dependency of $\mu_n$ on radius for smaller diameter NWs is expected. Since surface roughness scattering is nearly independent of temperature, the difference between the observed trends at 50K and 298K arise from a combination of phonon scattering, and surface/interface traps and fixed charges that contribute to additional surface scattering and lower gate coupling.

Future theoretical analysis of the various scattering events discussed above are needed to enable more detailed and quantitative understanding of the role of each scattering mechanism for a given NW radius and temperature range. Additionally, the electron effective mass may increase with the diameter miniaturization which could also have an impact in the diameter dependency of the mobility, and requires future theoretical insights. Clearly, the results presented here demonstrate the drastic effect of NW radius on the field-effect mobility. This is of concern since small diameter NWs ($r<\sim 10$ nm) are highly desirable for the channel material of future sub-10 nm FETs as they enable improved gate electrostatic control of the channel and lower leakage currents. However, this work suggests that the aggressive diameter scaling of NWs may only be attained at the cost of field-effect mobility degradation, therefore, requiring careful device design considerations for achieving the optimal device performances. Additionally, improving the surface properties is essential for enhancing the electron transport characteristics and the electrostatics of InAs NW-FETs.[10,12] A similar approach of utilizing *C-V* and *I-V* characterizations may be used in the future to systematically study the precise role of surface functionalization or high-κ gate dielectric integration on the electrical properties of InAs NW-FETs.

In summary, an approach for in-depth characterization of the intrinsic electronic properties of nanoscale materials is presented by utilizing detailed *C-V* and *I-V* measurements. Specifically, the *C-V* behavior of single InAs NW-FETs was successfully characterized for different temperatures and measuring frequencies. From the *C-V* measurements, information regarding $C_{ox}$, $C_{it}$ and $D_{it}$ was directly acquired while enabling the accurate assessment of field-effect mobility. The room temperature, field-effect mobility is found to linearly increase with radius for $r$=7-18 nm. The dependency of mobility on radius at low temperature (i.e., 50K) where the phonons and interface traps are thermally frozen out sheds light on the enhanced role of surface transport and surface scattering in smaller NWs. In the future, this approach can be utilized to systematically study the effects of surface passivation on the field-effect mobility and surface/interface traps and fixed charges.

**Acknowledgement**

This work was supported by Intel Corporation, MARCO/MSD Focus Center Research Program, and Berkeley Sensor and Actuator Center. J.C.H. acknowledges an Intel Graduate Fellowship. The synthesis part of this work was supported by a LDRD from Lawrence Berkeley National Laboratory. All fabrication was performed at the UC Berkeley Microlab facility. We thank Z.A. Jacobson for help with fabrication.

**Supporting Information Available:**

NW-FET fabrication details; *C-V* measurement set up; frequency-dependent *C-V* characterization; effective mobility; and the effect of quantum capacitance on the total gate capacitance. These materials are available free of charge via the Internet at http://pubs.acs.org.

**Figure Captions**

**Figure 1.** Electron microscopy characterization of InAs NWs. (a) SEM image of InAs NWs grown on a $SiO_2$/Si substrate by using Ni nanoparticles as the catalyst. (b) TEM image of a representative InAs NW. The inset shows the corresponding diffraction pattern converted by fast-Fourier transform where the zone axis of [110] can be identified. (c) The corresponding high resolution TEM image taken from the NW in (b). (d) The EDS analysis shows that the chemical composition of In:As is close to 1:1. (e) A top-view schematic of a global back-gated NW FET, used for the *I-V* characterization. (f) SEM image of a representative back-gated NW FET.

**Figure 2.** *I-V* characterization of InAs NW FETs. (a) Device output characteristics normalized for channel length ($I_{DS}.L$-$V_{GS}$) at $V_{DS}$=0.1 V for three separate long channel devices (*L*=8.4, 9.6, and 8.4 μm, respectively) with NW radii of *r*=17.5, 12.5, and 7.5 nm, respectively. The 2.5 nm oxide shell was subtracted from the measured NW radius. Length normalized $I_{DS}.L$-$V_{DS}$ plots for various $V_{GS}$ for the (b) 7.5 nm, (c) 12.5 nm, and (d) 17.5 nm radius NW devices. The NW diameter for each device was directly obtained from the atomic force microscopy and SEM analyses, with an uncertainty of ~±1 nm. All measurements were conducted in a vacuum ambient with minimal hysteresis.[21]

**Figure 3.** *C-V* characterizations of InAs NW-FETs. (a) Measurement schematics for *C-V* measurement of a single NW device (top); and the equivalent capacitance circuits in the depletion regime for low frequency (LF) and high frequency (HF) measurements (bottom). H and L represent the "high" and "low" terminals of the bridge, respectively. (b) Temperature dependent *C-V* characteristics for a local-gated NW FET with *r*~11 nm and $L_{LG}$~4.7 μm.

Electrostatic modeling is also applied and fitted to all measurements for the normalized gate capacitance.

**Figure 4.** Measured and simulated gate oxide capacitance as a function of radius per unit of local buried gate length. For the simulation a semiconductor nanowire with $\varepsilon=15$ was assumed. Additionally, the capacitance values obtained from the analytical expression of $C_{ox} = \dfrac{2\pi\varepsilon\varepsilon_0 L}{\cosh^{-1}[(r+t_{ox})/r]}$ are shown.

**Figure 5.** Room temperature mobility assessment. (a) Field-effect mobility as a function of $V_{GS}$ for three NWs of different radius ($r$=17.5, 12.5, and 7.5 nm), corresponding to the $I_{DS}.L$-$V_{GS}$ plot of Figure 2a. (b) Peak field-effect mobility as a function of radius for more than 50 different devices with NWs ranging from 7-18 nm in radius post oxide subtraction. Over this NW radius range, the peak field-effect mobility linearly increases with radius, closely fitting the linear expression $\mu_n$=422$r$-1180. Note that the $I_{DS}$-$V_{GS}$ plots were smoothed before the transconductance, $g_m$ was calculated for field-effect mobility assessment.

**Figure 6.** Temperature dependent electron transport properties. (a) $I_{DS}$-$V_{GS}$ at $V_{DS}$=0.01 V for a representative NW FET with $r$=18 nm and $L$=6.7 μm over a temperature range of 50-298 K. (b) The corresponding peak, field-effect mobility as a function of temperature for the same device. (c) The dependency of field-effect mobility on radius for four NWs of different radius at temperatures of 50 and 298 K.

# References

[1] Lieber, C. M.; Wang, Z. L. *MRS Bull.* **2007**, *32*, 99-104.

[2] Javey, A. *ACS Nano*. **2008**, *2*, 1329-1335.

[3] Bryllert, T.; Wernersson, L. E.; Froberg, L. E.; Samuelson, L. *IEEE Electron Device Lett.* **2006**, *27*, 323-325.

[4] Friedman, R. S.; McAlpine, M. C.; Ricketts, D. S.; Ham, D.; Lieber, C. M. *Nature* **2005**, *434*, 1085.

[5] Fan, Z.; Ho, J. C.; Jacobson, Z. A.; Razavi, H.; Javey, A. *Proc. Natl. Acad. Sci. U.S.A.* **2008**, *105*, 11066–11070.

[6] Fan, Z.; Ho, J. C.; Jacobson, Z. A.; Yerushalmi, R.; Alley, R. L.; Razavi, H.; Javey, A. *Nano Lett.* **2008**, *8*, 20-25.

[7] Wang, X. D.; Song, J. H.; Liu, J.; Wang, Z. L. *Science* **2007**, *316*, 102-105.

[8] Xiang, J.; Lu, W.; Hu, Y. J.; Wu, Y.; Yan, H.; Lieber, C. M. *Nature* **2006**, *441*, 489-493.

[9] McAlpine, M. C.; Ahmad, H.; Wang, D.; Heath, J. R. *Nature Mat.* **2007**, *6*, 379-384.

[10] Jiang, X.; Xiong, Q.; Nam, S.; Qian, F.; Li, Y.; Lieber, C. M. *Nano Lett.* **2007**, *7*, 3214-3218.

[11] Dayeh, S.; Soci, S.; Yu, P.; Yu, E.; Wang, D. *Appl. Phys. Lett*. **2007**, *90*, 162112-1-3.

[12] Hang, Q.; Wang, F.; Carpenter, P. D.; Zemlyanov, D.; Zakharov, D.; Stach, E. A.; Buhro, W. E.; Janes, D. B. *Nano Lett.* **2008**, *8*, 49-55.

[13] Lind, E.; Persson, A. I.; Samuelson, L.; Wernersson, L. E. *Nano Lett.* **2006,** *6,* 1842-1846.

[14] Bleszynski, A.C.; Zwanenburg, F. A.; Westervelt, R. M.; Roest, A. L.; Bakkers, E. P. A. M.; Kouwenhoven, L. P. *Nano Lett.* **2007**, *7*, 2559-2562.

[15] Lenzi, M.; Gnudi, A.; Reggiani, S.; Gnani, E.; Rudan, M.; Baccarani, G. *J. Comput. Electronics.* **2008**, *7*, 355-358.

[16] Kotlyar, R.; Obradovic, B.; Matagne, P.; Stettler, M.; Giles, M. D. *Appl. Phys. Lett.* **2004**, *84*, 5270-5272.

[17] Ramayya, E. B.; Vasileska, D.; Goodnick, S. M.; Knezevic, I. *IEEE Trans. Nanotechnology* **2007**, *6*, 113 – 117.

[18] Koo, S.-M.; Fujiwara, A.; Han, J.-P.; Vogel, E. M.; Richter, C. A.; Bonevich, J. E. *Nano Lett*. **2004**, *4*, 2197-2201.

[19] Chen, J.; Saraya, T.; Miyaji, K.; Shimizu, K.; Hiramoto, T. Symposium on VLSI Technology, **2008**, 32-33.

[20] Motayed, A.; Vaudin, M.; Davydov, A. V.; MeIngailis, J.; He, M.; Mohammad, S. N. *Appl. Phys. Lett.* **2007**, *90*, 043104.

[21] Ford, A. C.; Ho, J. C.; Fan, Z.; Ergen, O.; Altoe, V.; Aloni, S.; Razavi, H.; Javey, A. *Nano Res*. **2008**, *1*, 32-39.

[22] Chueh, Y.-L.; Ford, A. C.; Ho, J. C.; Jacobson, Z. A.; Fan, Z.; Chen, C.-Y.; Chou, L.-J.; Javey, A. *Nano Lett*., **2008**, ASAP.

[23] Karlström, O.; Wacker, A.; Nilsson, K.; Astromskas, G.; Roddaro, S.; Samuelson, L.; Wernersson, L. E. *Nanotech*. **2008**, *19*, 435201.

[24] Ilani, S.; Donev, L. A. K.; Kindermann, M.; McEuen, P. L. *Nature Phys.* **2006**, *2*, 687-691.

[25] Tu, R.; Zhang, L.; Nishi, Y; Dai, H. *Nano Lett.* **2007**, *7*, 1561-1565.

[26] Javey, A.; Kim, H.; Brink, M.; Wang, Q.; Ural, A.; Guo, J.; McIntyre, P.; McEuen, P.; Lundstrom, M.; Dai, H. *Nature Mat*. **2002**, *1*, 241-246.

[27] Wang, D.; Wang, Q.; Javey, A.; Tu, R.; Dai, H.; Kim, H.; Krishnamohan, T.; McIntyre, P.; Saraswat, K. *Appl. Phys. Lett*. **2003**, *83*, 2432-2434.

[28] Duan, X.; Huang, Y.; Cui, Y.; Wang, J.; Lieber, C. M. *Nature* **2001**, *409*, 66-69.

[29] Huang, Y,; Duan, X.; Cui, Y.; Lieber, C. M. *Nano Lett*. **2002**, *2*, 101-104.

[30] Bryllert, T.; Samuelson, L.; Jensen, L.; Wernersson, L. *DRC Proc*. **2005**, *1*, 157.

[31] Guo, J. unpublished, **2008**.

[32] Leonard, F.; Talin, A. *Phys. Rev. Lett*. **2006**, *97*, 026804-1-4.

[33] Arora, N. *MOSFET Modeling for VLSI Simulation.* World Scientific, **2006**.

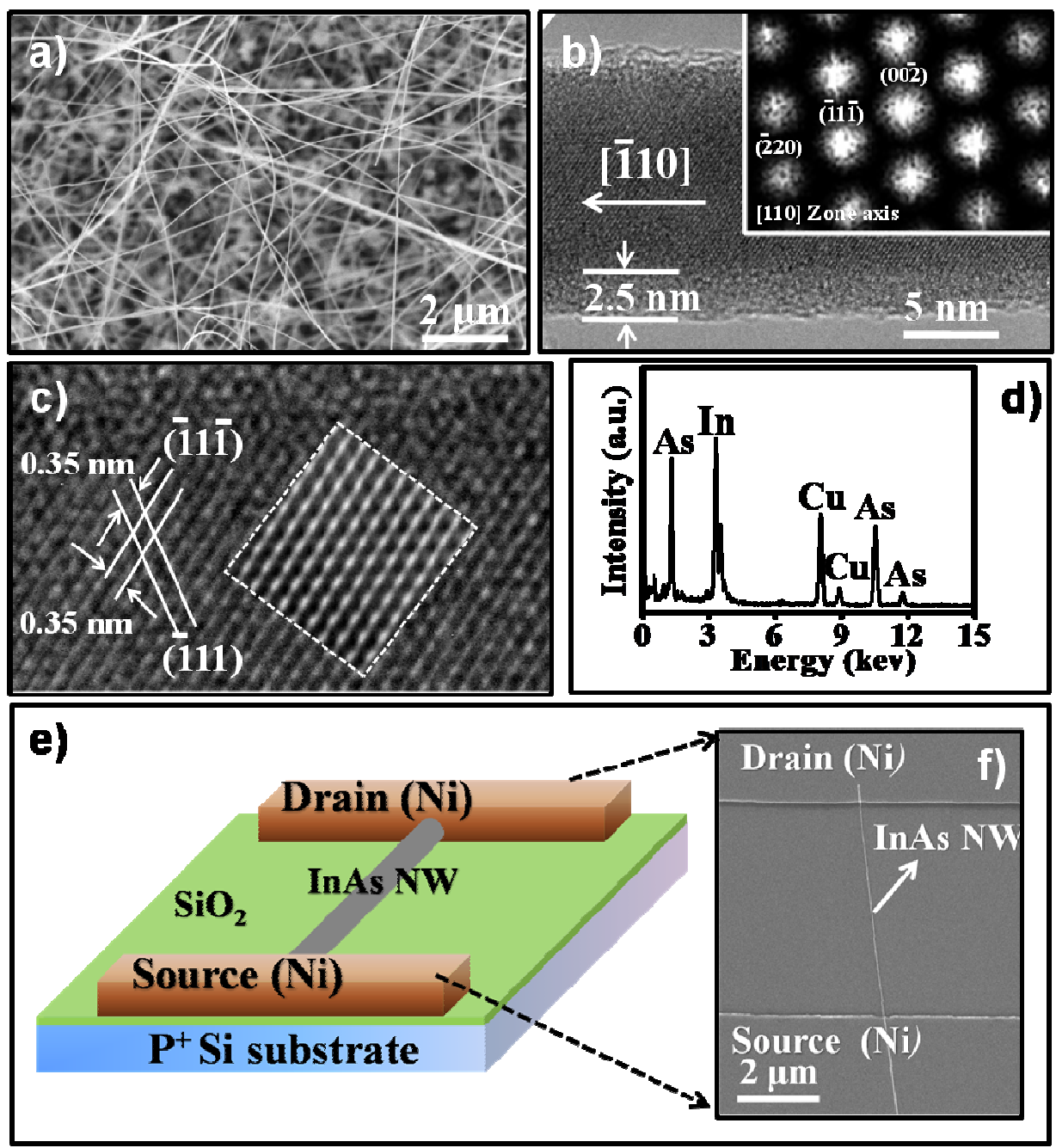

**Figure 1**

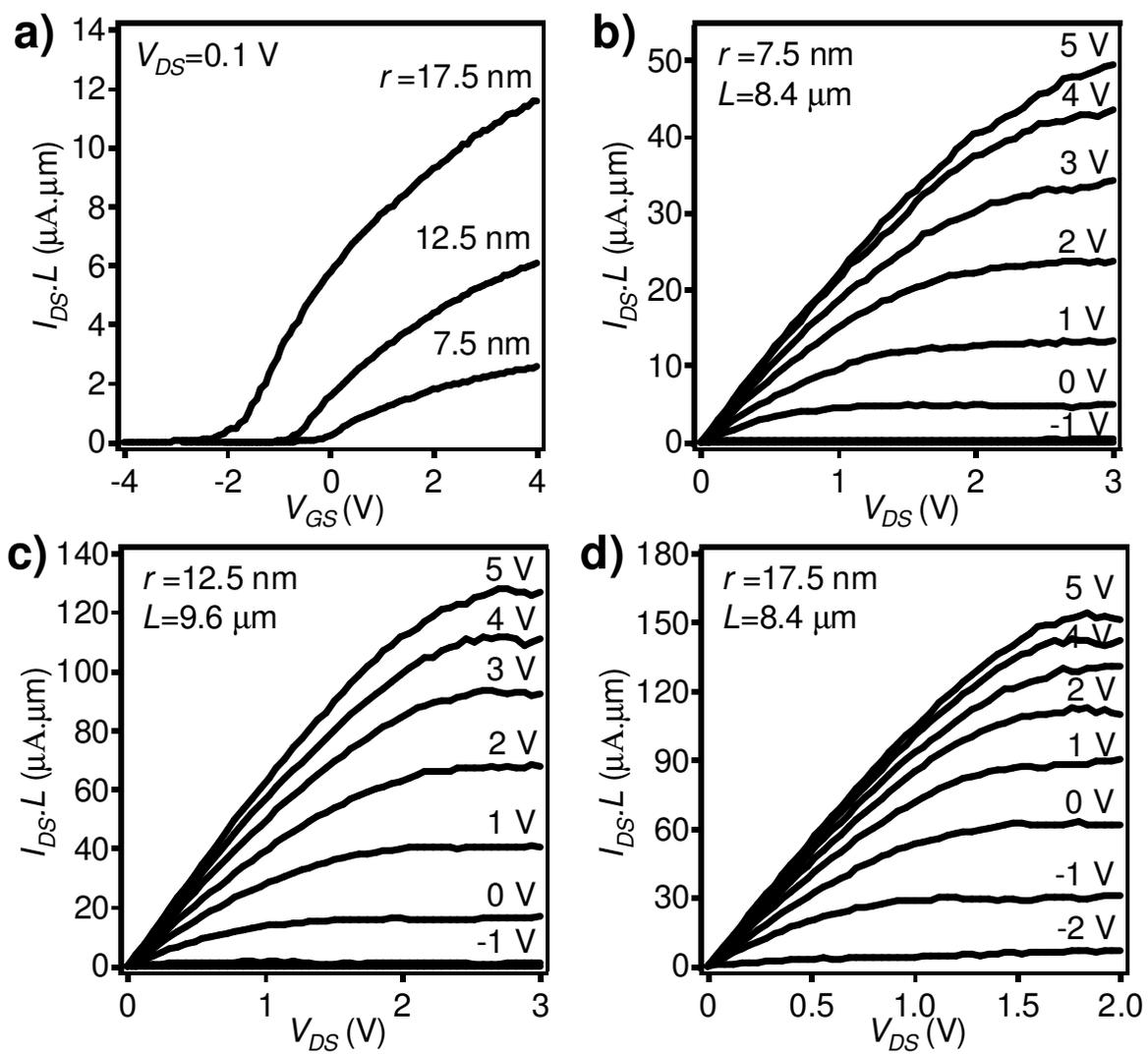

**Figure 2**

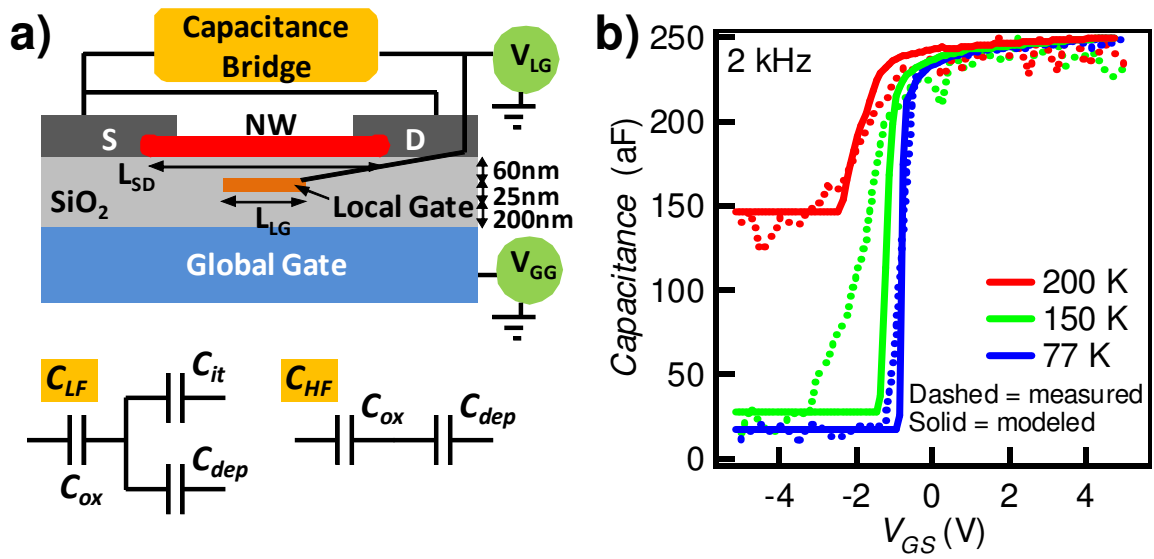

**Figure 3**

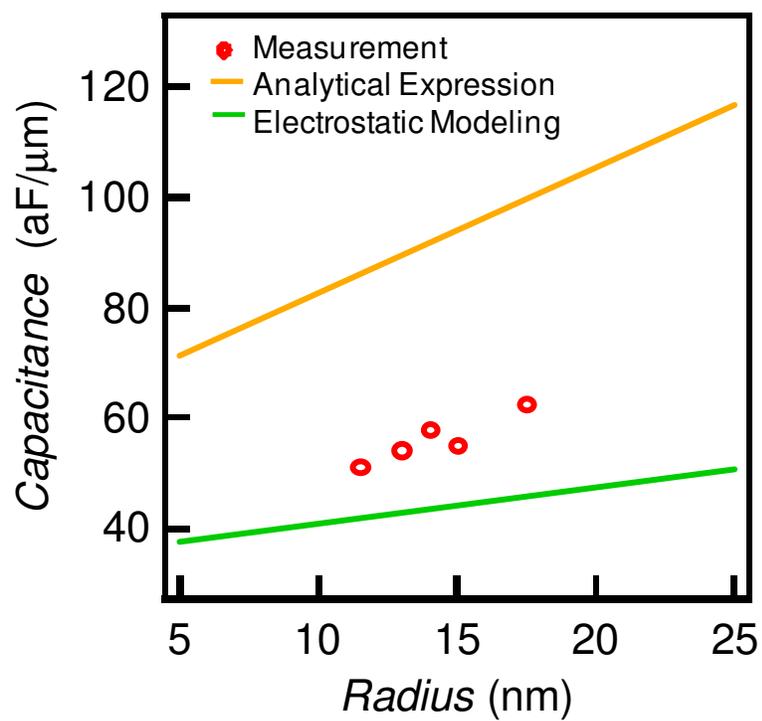

**Figure 4**

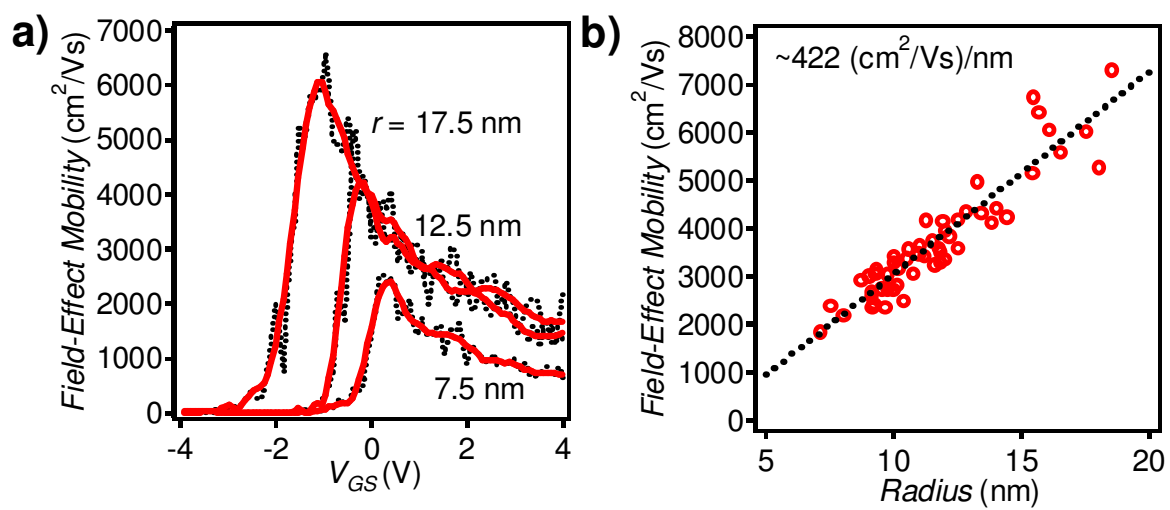

**Figure 5**

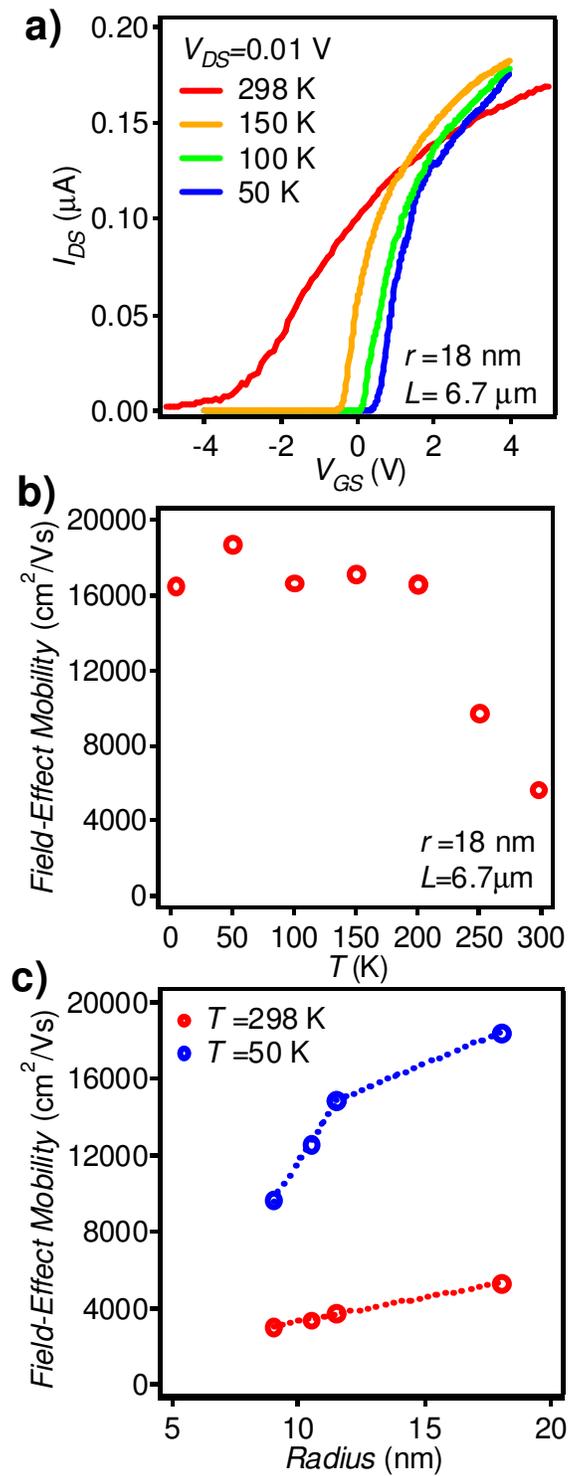

**Figure 6**

# TOC

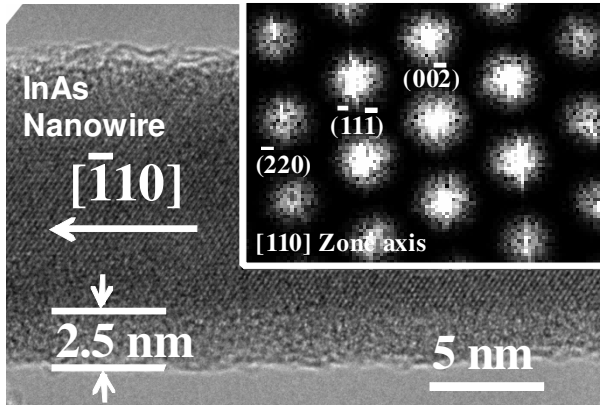 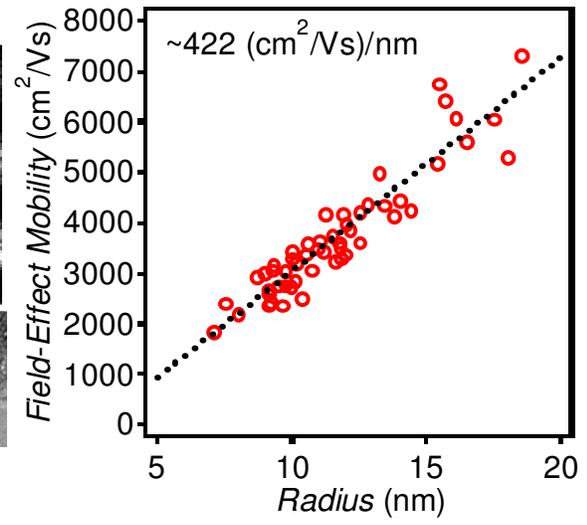